\PassOptionsToPackage{unicode}{hyperref}
\PassOptionsToPackage{hyphens}{url}
\documentclass[
]{article}
\usepackage{amsmath,amssymb}
\usepackage{lmodern}
\usepackage{iftex}
\ifPDFTeX
  \usepackage[T1]{fontenc}
  \usepackage[utf8]{inputenc}
  \usepackage{textcomp} 
\else 
  \usepackage{unicode-math}
  \defaultfontfeatures{Scale=MatchLowercase}
  \defaultfontfeatures[\rmfamily]{Ligatures=TeX,Scale=1}
\fi
\IfFileExists{upquote.sty}{\usepackage{upquote}}{}
\IfFileExists{microtype.sty}{
  \usepackage[]{microtype}
  \UseMicrotypeSet[protrusion]{basicmath} 
}{}
\makeatletter
\@ifundefined{KOMAClassName}{
  \IfFileExists{parskip.sty}{%
    \usepackage{parskip}
  }{
    \setlength{\parindent}{0pt}
    \setlength{\parskip}{6pt plus 2pt minus 1pt}}
}{
  \KOMAoptions{parskip=half}}
\makeatother
\usepackage{xcolor}
\IfFileExists{xurl.sty}{\usepackage{xurl}}{} 
\usepackage{hyperref}
\hypersetup{
  pdftitle={Against Interaction Design},
  pdfauthor={Tim Murray-Browne},
  hidelinks,
  pdfcreator={LaTeX via pandoc}}
\usepackage{graphicx}
\makeatletter
\def\maxwidth{\ifdim\Gin@nat@width>\linewidth\linewidth\else\Gin@nat@width\fi}
\def\maxheight{\ifdim\Gin@nat@height>\textheight\textheight\else\Gin@nat@height\fi}
\makeatother
\setkeys{Gin}{width=\maxwidth,height=\maxheight,keepaspectratio}
\makeatletter
\def\fps@figure{htbp}
\makeatother
\ifLuaTeX
  \usepackage{selnolig}  
\fi

\title{Against Interaction Design}
\author{Tim Murray-Browne}
\date{2022-09-30}

\usepackage[square,sort,comma,numbers]{natbib}

\begin{document}
\maketitle

{\it

\textbf{Against Interaction Design} is a short manifesto that distils a
position that's emerged through a decade of creating interactive art.

I intend it here as a provocation and a speculation on an alternative
future relationship between people and machines.

}

\begin{center}\rule{0.5\linewidth}{0.5pt}\end{center}

\section{Against Interaction Design}

\subsection{To design interaction is to design who we can
be}\label{to-design-interaction-is-to-design-who-we-can-be}

Look at the words on my keyboard: command, control,
function, shift. \textbf{These describe the processes of a
factory or a military unit}, not a conversation, nor a dance, nor
friends eating together. What of the rest of being human?

Behind every interface is a model world. People are modelled as
profiles. Emotions are modelled as emojis. The interface defines how I
can sense and shape that world. It defines my relationship to that
world: what I can do, who I am, and who I can be.

To interact \emph{through} an interface, I need to think in terms of the
model of that interface. I need to translate my vague, complex
intentions into the language of that interface. I need to align my
thoughts and actions to the structures and processes conceived by its
designer. I need to see myself as a part of the model the interface
implies.

The more my interaction is designed, the more involved the designer
becomes with not just what I can do, but how and why I might do it. I
need to be the model user the designer imagines me to be. And so the
less space I have to express the nuance of who I really am.

An `expert` user is someone who transforms into the person the interface
needs them to be. Those who struggle or refuse to adapt are excluded.

This is not so much a problem of bad interaction design as it is a
problem of what interaction design is.

\subsection{The asymmetry of interaction
design}\label{the-asymmetry-of-interaction-design}

Marshall McLuhan argued that we shape our tools, and thereafter they
shape us. Except, it is not us who shape our interfaces. Interaction
designers do.

We all use the same few interfaces crafted by a handful of people. Each
design decision, however heroic, however thoughtful, however detailed,
affects us all in the same way. The more interaction is designed, the
more human agency becomes constrained to fit within the models conceived
by the designer.

\textbf{The asymmetry of interaction design gives interaction designers
too much leverage over our lives.}

When a single interface becomes standardised across so much of our
lives, it makes us all share its singular model of that world. This
works for standardised tasks like flying a plane or filing a tax return.
It works for a game where we can freely step in and out of the role it
defines for us.

But it doesn't work for human expression and social connection because
\textbf{it filters out, rather than amplifying, the chaos that makes us
human}.

\subsection{The standardisation of
self-expression}\label{the-standardisation-of-self-expression}

Its singular model centres what matters and sidelines what doesn't. It
replaces diversity with efficiency. It reduces our messy needs into
explicit goals we can all agree on.

But the rest of \textbf{being human does not reduce}. To express what we
think, we must continually adopt and discard different models of the
world. A model is always an imperfect approximation. But what we are
trying to express is real. There is no single model that can capture it
all.

A standard interface that spreads to define how we relate to each other,
to society, to ideas, our ability to express who we are? This is
dangerous.

A standard interface standardises how we think. Some of the worst deeds
of humanity emerge when the plethora of human wildness is filtered
through a singular model of the world. Singular models become invisible.
We forget how to discard them. Monoculture.

\emph{`If all you have is a hammer, everything looks like a
nail.'}\footnote{This quote is often attributed apocryphally to
  \href{https://en.wikipedia.org/wiki/Law_of_the_instrument}{Abraham
  Maslow}.}

If all anyone has is a hammer, we all become nails.

\subsection{Interaction without Interaction
Design}\label{interaction-without-interaction-design}

When it comes to human expressivity, we cannot free people by designing
how they can behave. But we can help people free themselves by letting
them negotiate their own relationship with technology.

We need interfaces that flex, adapt and warp as needed by the human,
\textbf{interfaces that emerge through negotiation between human and
system}. We need interfaces as diverse as the people that connect
through them.

An alternative to Designed Interaction is \textbf{Negotiated
Interaction}.

Negotiated Interaction aims to uncover interfaces that:

\begin{itemize}

\item Let us interact without being subjected to Interaction Design.

\item Belong to the individual as much as they do to the technological system.

\item Empower us without making us translate our needs, thoughts and impulses into the worldview of its designer.

\item Let us express ourselves through technology with as much individuality, vulnerability and sensitivity as we have without technology.

\item Let us communicate our empathetic, emotional and embodied selves through technology without the need for explanation, rationalisation or
categorisation.

\item Let us create digital forms with as much ambiguity and complexity as our lived experiences.
\end{itemize}

\textbf{Negotiated Interaction supplants command and control with
resonate and unite.}

\begin{center}\rule{0.5\linewidth}{0.5pt}\end{center}

\section{Epilogue}\label{epilogue}

To see our practical explorations of Negotiated Interfaces, see Sonified
Body project\footnote{\url{https://timmb.com/sonified-body}} (2021, collaboration with Panagiotis Tigas), which uses AI
to generate bespoke interfaces based on a dancer's existing vocabulary
of movement.

Also see Latent Voyage\footnote{\url{https://timmb.com/latent-voyage}}, which applies this technique to generate an
embodied, non-symbolic interface for an untrained person to interact
with an image-generating AI.

For a more academic treatment, see our peer-reviewed paper \emph{Emergent
Interfaces: Vague, Complex, Bespoke and Embodied Interaction between
Humans and Computers} \cite{murray-browne2021emergent-interfaces}.

These ideas have been fueled by the writings of many others. See \cite{mcgilchrist2009master-emissary,mcneil2020lurking, lanier2018ten-arguments, zuboff2019surveillance-capitalism,dourish2004where-the-action-is,rokeby1998construction-of-experience} for suggestions of further reading, and \cite{escobar2018designs-for-the-pluriverse} for a suggestion from a reader of an earlier draft of this text.

\section{Acknowledgements}

Thank you to Adriana Minu, Matt Thompson, Panagiotis Tigas, Tadeo Sendon
and Nela Cicmil for offering feedback on an earlier draft of this text.

\bibliography{2022-05-20_Against_interaction_design.bib}

\end{document}